\documentclass[aip,apl,amsmath,amssymb,reprint,preprintnumbers, floatfix]{revtex4-1}

\usepackage{graphicx}
\usepackage{dcolumn}
\usepackage{bm}
\usepackage{upgreek} 
\usepackage{url}

\begin{document}

\preprint{AIP/123-QED}

\title[Sample title]{Ga$_{1-x}$Mn$_{x}$N epitaxial films with high magnetization}

\author{G. Kunert}
 \email{kunert@ifp.uni-bremen.de}
\affiliation{Semiconductor Epitaxy, Institute of Solid State Physics, University of Bremen, D-28359 Bremen, Germany}
\author{S. Dobkowska}
\affiliation{Institute of Physics, Polish Academy of Science, PL-02-668 Warszawa, Poland}
\author{Tian Li}
\affiliation{Institute of Solid State Physics, Johannes Kepler University Linz, A-4040 Linz, Austria}
\author{H. Reuther}
\affiliation{Helmholtz-Zentrum Dresden-Rossendorf, Institute of Ion Beam Physics and Materials Research, Bautzner Landstra\ss{}e 400, 01328 Dresden}
\author{C. Kruse}
\affiliation{Semiconductor Epitaxy, Institute of Solid State Physics, University of Bremen, D-28359 Bremen, Germany}
\author{S. Figge}
\affiliation{Semiconductor Epitaxy, Institute of Solid State Physics, University of Bremen, D-28359 Bremen, Germany}
\author{R. Jakiela}
\affiliation{Institute of Physics, Polish Academy of Science, PL-02-668 Warszawa, Poland}
\author{A. Bonanni}
\affiliation{Institute of Solid State Physics, Johannes Kepler University Linz, A-4040 Linz, Austria}
\author{J. Grenzer}
\affiliation{Helmholtz-Zentrum Dresden-Rossendorf, Institute of Ion Beam Physics and Materials Research, Bautzner Landstra\ss{}e 400, 01328 Dresden}
\author{W. Stefanowicz}
\affiliation{Institute of Physics, Polish Academy of Science, PL-02-668 Warszawa, Poland}
\author{M. Sawicki}
\affiliation{Institute of Physics, Polish Academy of Science, PL-02-668 Warszawa, Poland}
\author{T. Dietl}
\affiliation{Institute of Physics, Polish Academy of Science, PL-02-668 Warszawa, Poland}
\affiliation{Institute of Theoretical Physics, Faculty of Physics, University of Warsaw, PL-00-681 Warszawa, Poland}
\author{D. Hommel}
\affiliation{Semiconductor Epitaxy, Institute of Solid State Physics, University of Bremen, D-28359 Bremen, Germany}

\date{\today}

\begin{abstract}
We report on the fabrication of pseudomorphic wurtzite Ga$_{−x}$Mn$_x$N grown on GaN with Mn concentrations up to 10\,\% using molecular beam epitaxy. According to Rutherford backscattering the Mn ions are mainly at the Ga-substitutional positions, and they are homogeneously distributed according to depth-resolved Auger-electron spectroscopy and secondary-ion mass-spectroscopy measurements. A random Mn distribution is indicated by transmission electron microscopy, no Mn-rich clusters are present for optimized growth conditions. A linear increase of the $c$-lattice parameter with increasing Mn concentration is found using x-ray diffraction. The ferromagnetic behavior is confirmed by superconducting quantum-interference measurements showing saturation magnetizations of up to 150\,emu/cm$^{3}$.

\end{abstract}

\pacs{75.50.Pp, 75.60.Ej, 81.05.Ea, 81.15.Hi}
\keywords{GaN, GaMnN, DMS}
\maketitle

Dilute magnetic semiconductors (DMSs) \cite{ohno_window_2010, dietl_ten-year_2010} and particularly Mn-based nitrides, have been in the focus of research interest since the prediction that these systems may show hole-mediated room temperature ferromagnetism, if challenges associated with solubility limits, self-compensation, and the transition to a strong p-d coupling case are overcome.\cite{Dietl} The potential of this type of materials to combine traditional electronics and photonics with the spin degree of freedom triggered the development of a number of potential applications.\cite{ohno_window_2010, Bader_Spintronics} Especially Ga$_{1-x}$Mn$_x$N is believed to be able to meet the conditions necessary for the implementation in devices, though the Mn solubility and the presence of compensating defects depend crucially on the growth conditions.\cite{Bonanni_transition_2007} It is now well established that because of the strong p-d hybridization, the Mn$^{2+/3+}$ acceptor level occupies the 
mid band gap position in GaN.\cite{graf_prospects_2003, wolos_configuration_2009} Accordingly, in samples containing compensating donor centers, Mn ions assume the 2+ 
charge state, for which short range antiferromagnetic interactions dominate.\cite{zajac_paramagnetism_2001, granville_neighbour_2010, Bonanni_PRB_2011} In contrast, in uncompensated films, where Mn$^{3+}$ ions prevail,\cite{Bonanni_PRB_2011, Sarigannidou_PRB_2006, Freeman_PRB_2007, Stefanowizc_PRB_2010} the net superexchange
interactions become ferromagnetic, \cite{Bonanni_PRB_2011, Sarigannidou_PRB_2006, Kondo_JCG_2002, Sawicki_arXiv_2012}  as predicted for Cr$^{2+}$ in II-VI DMSs,\cite{Blinowski_PRB_1996} leading to Curie temperatures up to 8\,K for $x\,\approx$\,6\,\%.\cite{Bonanni_PRB_2011, Sawicki_arXiv_2012} It has been suggested that the presence of ferromagnetic interactions without band carriers, together with a sizable spin splitting of the excitonic states indicates the suitability of this dilute ferromagnetic insulators for magnetooptical devices, such as optical isolators. This has been already revealed for Ga$_{1-x}$Mn$_x$N,\cite{PRB_Suffczynski_2011} where the destructive effect of antiferromagnetic interactions specific to II-VI Mn-based DMSs is circumvented.\cite{Zayets_JOptSoc_2005}

In this letter, we report on the growth conditions that allow to obtain pseudomorph Ga$_{1-x}$Mn$_x$N films containing up to 10\,\% of randomly distributed Mn ions using molecular beam epitaxy. In order to clarify the structural and magnetic properties of these samples, several investigation methods have been employed. The $c$-axis lattice parameter of these films is found to obey Vegard's law. Furthermore, the observed values of saturation magnetization up to 150\,emu/cm$^3$ are superior to other DMSs, such as (Ga,Mn)As.\cite{ohya_APL_2007, chiba_APL_2007, mack_APL_2008, wang_APL_2008} 

An EPI 930 molecular beam epitaxy (MBE) chamber equipped with a radio-frequency plasma source is used to deposit the Ga$_{1-x}$Mn$_x$N layers. Templates consisting of 2\,$\upmu$m GaN(0001) on c-plane sapphire fabricated by metal-organic vapor-phase epitaxy are used as substrates, the MBE grown layers are about 200\,nm thick. All samples dicussed here are grown at a substrate temperature of 760\,$^\circ$C. At this temperature layer-by-layer growth is observed for metal-rich growth conditions. Mn acts as a surfactant for the growth, while excess material evaporates due to the comparatively low sticking coefficient of this element on GaN. In order to hinder the incorporation of Mn clusters within the layers, excessively metal-rich growth conditions are avoided during epitaxial deposition by carefully adjusting the Ga and Mn flux. Slightly metal-rich growth conditions close to the transition from two-dimensional to three-dimensional growth result in smooth layers without additional crystalline phases, while clusters are observed for highly Mn-rich conditions. A nitrogen flux of 1.3 standard cubic centimeters per minute at a plasma power of 300\,W is used to provide atomic nitrogen.  For each sample a beam equivalent pressure (BEP) of the Mn between 8$\times$10$^{-8}$ and 2.1$\times$10$^{-7}$\,Torr was chosen, while the flux of the Ga cell is reduced from 2$\times$10$^{-6}$\,Torr to 1.15$\times$10$^{-7}$\,Torr BEP. As the result, Ga$_{1-x}$Mn$_x$N layers with Mn contents up to 10\,\% are obtained.
\begin{figure}[htb]
\includegraphics[width=\linewidth]{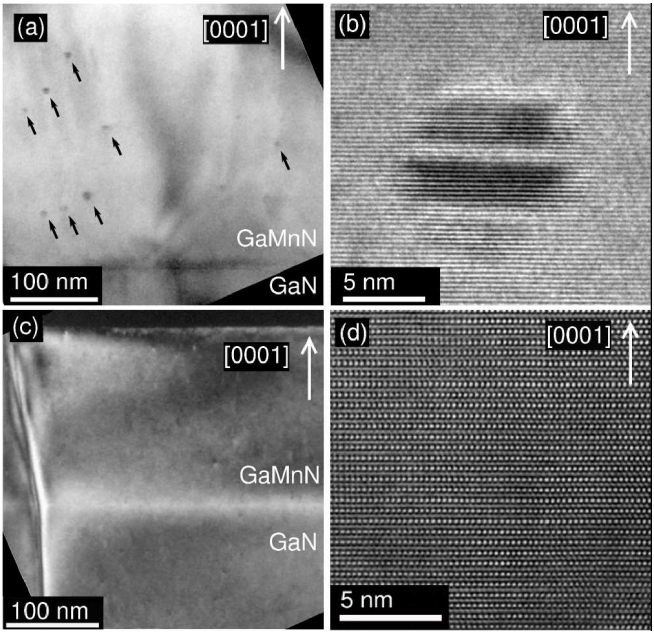}
\caption{\label{fig:TEM} Low magnification TEM (a,c) and high magnification TEM (b,d) for Ga$_{1-x}$Mn$_x$N layers. Secondary crystallographic phases, marked by black arrows, are observed as Moir\'{e} fringes for Mn rich growth conditions (a,b). No evidence of crystallographic phase separation is detected for growth conditions close to the stoichiometric point (c,d).}
\end{figure}

Transmission electron microscopy investigations have been carried out using a JEOL 2011 Fast TEM microscope operating at 200\,kV. The samples are prepared by mechanical polishing. Afterwards ion polishing is used to flatten the surface of the specimen. Images in different magnification are depicted in Fig.~\ref{fig:TEM}. The sample shown in Fig.~\ref{fig:TEM}(a) and (b) is grown under highly Mn-rich growth conditions. Crystallographic phase separation is detected as Moir\'{e} fringes. In contrast, no indication of phase separation or cluster incorporation is found for the Ga$_{1-x}$Mn$_x$N layers grown under slightly metal-rich conditions, as exemplarily shown in in Fig.~\ref{fig:TEM}(c) and (d), respectively. Mn is incorporated in a dilute manner solely in the wurtzite Ga$_{1-x}$Mn$_x$N phase. Furthermore, no indication of an additional formation of threading dislocations was found in these layers. In the following, only the samples without phase separation will be discussed. 

\begin{figure}[htb]
\includegraphics[width=\linewidth]{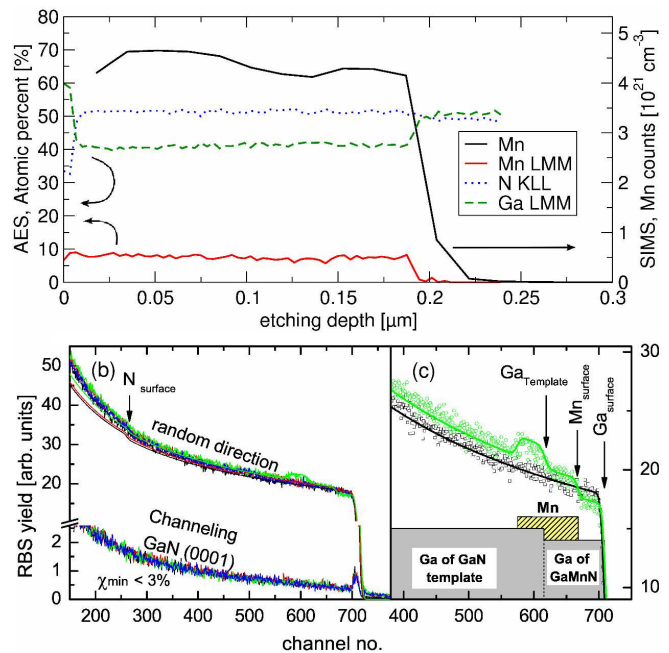}
\caption{\label{fig:Auger} (a) Spatially resolved SIMS profile and Auger electron spectroscopy scans for a Ga$_{1-x}$Mn$_x$N layer containing about 10\,\% Mn. They show a homogeneous distribution of Mn through the whole layer. Within the layer the Ga fraction is reduced according to the amount of Mn atoms incorporated. (b) RBS measurements are shown for GaN and Ga$_{1-x}$Mn$_x$N containing up to 10\,\% Mn. The in RBS yield drops to \textless3\,\% of the random signal in channeling mode and becomes indistinguishable. (c) Magnified random signals of the sample containing about 10\,\% Mn (green) compared to GaN (black). The simulated curves are represented by solid lines.}
\end{figure}

Auger electron spectroscopy (AES) as well as secondary ion mass spectrometry (SIMS) measurements have been carried out on selected samples to investigate the depth profile of Mn. In both methods Ar$^+$ ions are used for sputtering. In SIMS the secondary ions are detected directly by a mass spectrometer. For AES the KLL transition is used to detect nitrogen, while the atomic fraction of Ga and Mn is determined by quantifying the LMM transition, respectively. Both curves shown in Fig.~\ref{fig:Auger} reveal homogeneous elemental profiles for a typical sample containing a Mn content of about 10\,\%. The amount of Mn and Ga adds up to 50\,\% of the total number of atoms in a complementary manner, indicating Mn incorporation on Ga sites. The interface between Ga$_{1-x}$Mn$_x$N layer and the GaN template layer at 0.2\,$\upmu$m can clearly be identified as a step in the elemental concentrations in the depth profile. At this step the total number of Mn atoms per cubic centimeter drops five orders of magnitude to the noise level.

\begin{figure*}[t!]
\includegraphics[width=1.0\linewidth]{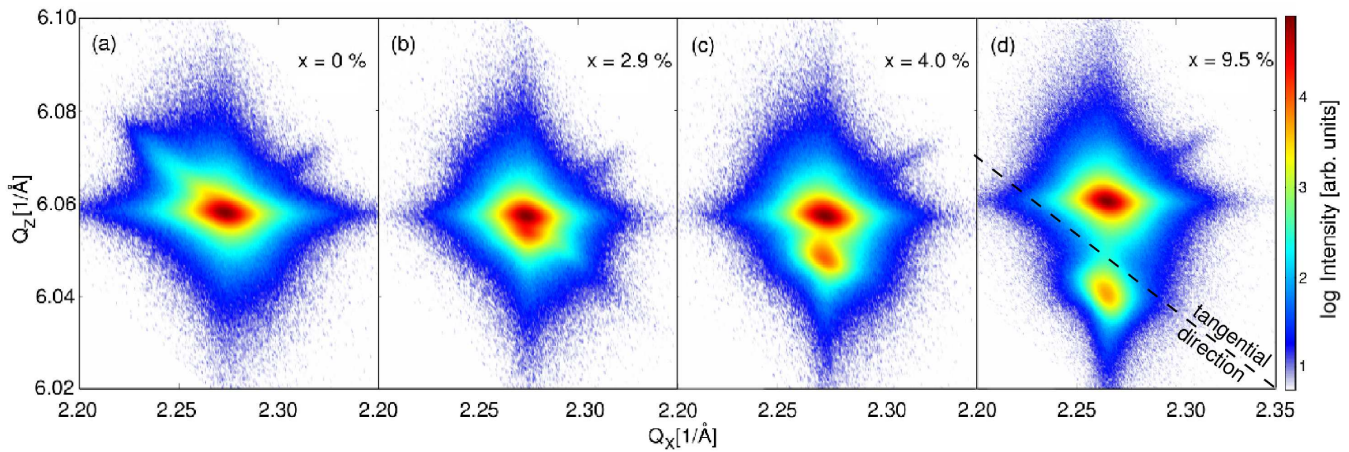}
\caption{\label{fig:RSM}Reciprocal space maps of the asymetric (10$\overline{1}$5) x-ray reflex. The high intensity peak is related to the GaN template layer. The low intensity peak origins from the Ga$_{1-x}$Mn$_x$N layer. Its position shifts with Mn content.}
\end{figure*}
In order to clarify whether Mn atoms are also incorporated on interstitial lattice sites to a certain extent, Rutherford backscattering (RBS) measurements have been carried out. He atoms with energies of 1.7\,MeV and full width at half maximum of 18\,keV are accelerated in random and (0001) channeling direction. Samples with Mn concentrations from 0\,\% (pure GaN) to about 10\,\% are investigated in channeling mode as well as in random mode [Fig.~\ref{fig:Auger}(b)].

In channeling mode the channeling signal drops to approximately 3\,\% of the random signal; the samples are indistinguishable one from another indicating the high crystalline quality. These measurements confirm the low defect density in the samples. There are almost no interstitials detected in the (0001) channels. The amount of Mn incorporated has been determined by simulating the measured signal using an appropriate model, dividing the investigated sample into sublayers and calculating the energy spectrum of the backscattered He ions using the program code SIMNRA.\cite{website_simnra} A magnification of the simulated elemental peaks is depicted in Fig.~\ref{fig:Auger}(c), exemplatory for the sample containing about 10\,\% Mn and pure GaN.

In order to verify a modification of the GaN lattice parameter induced by the Mn incorporation, high resolution x-ray diffraction (HRXRD) maps around the asymmetric (10$\overline{1}$5)-reflex are performed as depicted in Fig.~\ref{fig:RSM}. These show two distinct features: the high intensity peaks at $Q_z$\,=\,6.06\,\AA{}$^{-1}$ are related to the GaN template layer while the lower intensity peak at smaller reciprocal lattice units origins from the thinner Ga$_{1-x}$Mn$_x$N layer. The position of the peak shifts with increasing Mn content to smaller values of $Q_z$, \textit{i.e.} larger lattice parameter. The peak of the Ga$_{1-x}$Mn$_x$N layer has the same Q$_x$ position as the GaN template material, indicating a pseudomorphic growth. The orientation and shape of the Ga$_{1-x}$Mn$_x$N layer peak changes with increasing Mn content. This behavior could be attributed to a change of the microstructure. At low Mn contents (Fig.~\ref{fig:RSM}b, $x\approx$2.9\,\%) it is more a columnar like structure whereas at high Mn contents (Fig.~\ref{fig:RSM}d $x\approx$10\,\%) the Ga$_{1-x}$Mn$_x$N layer shows a  strong mosaicity effect as this peak is mainly elongated along the tangential direction.\cite{Heinke} Peak positions of 2$\theta-\omega$ scans of the (0002)- and the (10$\overline{1}$5)-reflex are used to determine the $c$-lattice parameter for layers containing different Mn concentrations (not shown).

\begin{figure}[htb]
\includegraphics[width=1.0\linewidth]{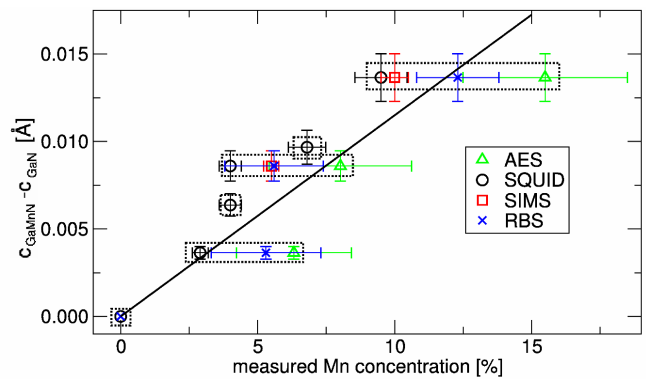}
\caption{\label{fig:vegard} Change of $c$-lattice parameter with Mn content, determined by different methods. Measurements belonging to the same sample are marked by a dashed rectangle.}
\end{figure}
In Fig.~\ref{fig:vegard}, the deviation of the $c$-lattice parameter of the Ga$_{1-x}$Mn$_x$N-layer from the GaN template is plotted against the Mn content determined by the four different methods RBS, AES, SIMS and magnetometry employing a superconducting quantum interference measurements device (SQUID). Dashed rectangles indicate measurements belonging to the same sample. The vertical error bars represent the standard distribution of the spacing measured on different positions of the sample and comprises the error due to lateral inhomogeneities caused by a temperature gradient on the sample during growth. Vegard's law describes a linear variation of lattice parameters with the composition of a mixed crystal (in relaxed state):%
\begin{equation}
c(x)=c_0+\Delta c\cdot x\hspace{1cm}a(x)=a_0+\Delta a\cdot x
\end{equation}
As the sampels are pseudomorphic the $c$-lattice parameter is additionally elongated due to the elastic distortion of the unit cell. Taking all measurements without weighing into account we obtain a slope of 0.115(9)\,$\mathrm{\AA{}}/x$ from a linear fit, where $x$ denotes the atomic Mn concentration as fraction of the total number of cation sites. Therefore, this formula can be used to determine the Mn content for Ga$_{1-x}$Mn$_x$N samples in a convenient way using the non-destructive XRD method alone. For nano-wires an increase of the $c$ and $a$-lattice parameter of $0.074\,\mathrm{\AA}/x$ and $0.048\,\mathrm{\AA}/x$, respectively was calculated.\cite{GaMnN_nanowires} Taking a ratio of the elastic constants of $2C_{13}/C_{33}\approx 0.5$ into account, these parameters result into an expected slope of $0.11\,\mathrm{\AA}/x$ for pseudomorphic material which is identical to our results within the error limits.\cite{Schuster_Strain_1999} For polycrystalline material a value of $m_{\text{poly}}$\,=\,0.53\,\AA{}/$x_{\text{Mn}}$ can be calculated from reference \onlinecite{leite_nanocrystalline_2006} in the same manner, which deviates significantly.
\begin{figure}[htb]
\includegraphics[width=0.85\linewidth]{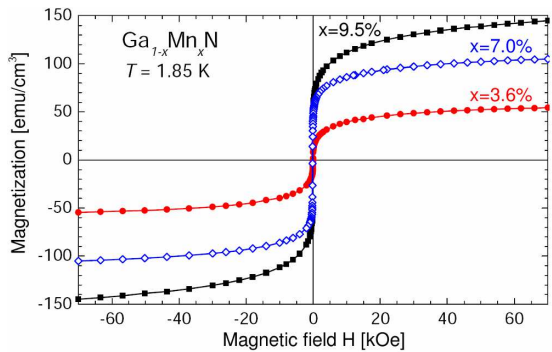}
\caption{\label{fig:SQUID} Magnetization curves measured at 1.85\,K for different Mn concentration $x$, evaluated from saturation values of magnetization. The external magnetic field is perpendicular to the c-axis of the wurtzite Ga$_{1-x}$Mn$_x$N crystal.}
\end{figure}

Magnetization measurements have been carried out down to a temperature of 1.85\,K employing the commercial SQUID magnetometer MPMS XL–7 from Quantum Design and a measurement technique specifically developed in order to meaningfully examine thin layers of magnetically dilute semiconductors.\cite{Sawicki:2011_SST} In Fig. \ref{fig:SQUID} magnetization values measured up to 70\,kOe for different Mn concentrations are depicted. While Curie temperatures are below 15\,K in these samples, they show sizable saturation magnetizations of up to 150\,emu/cm$^3$ for the highest concentrations achieved. Such values are the highest ever reported for DMS, exceeding by 50\,\% those for (Ga,Mn)As.\cite{chiba_APL_2007} The detailed structural analysis presented before allows us to rule out a statistically significant contribution from secondary magnetic phases, pointing to Ga-substitutional Mn diluted in GaN as the sole source of this magnetization. The Mn concentration is determined by the near-saturation value of the
magnetization at 70\,kOe and 1.85\,K, assuming that the value of the magnetic moment per Mn cation is 3.72\,$\upmu_B$ at these conditions, as implied by the group theoretical model for non-interacting Mn$^{3+}$ cations.\cite{ Stefanowizc_PRB_2010, Gosk_PRB_2005} As shown in Fig. \ref{fig:vegard}, the values of $x$ obtained in this way are in a good agreement with other determinations of the Mn content.

In summary, we have successfully fabricated epitaxially dilute Ga$_{1-x}$Mn$_x$N films. Extensive structural characterization excludes the presence of secondary crystalline phases or interstitial Mn atoms in samples obtained under appropriate growth conditions. A systematic modification of the $c$-lattice parameter is observed with increasing Mn content, which is used to determine the parameters for Vegard's law for this material for the pseudomorphic case. Magnetizations of up to 150\,emu/cm$^3$ are the highest reported for any DMS.

The authors would like to thank Dr. J. v. Borany for carrying out RBS measurements. This work was supported by the FunDMS Advanced Grant of the ERC (Grant No. 227690) within the Ideas 7th Framework Programme of European Community, by the InTechFun (Grant No. POIG.01.03.01-00-159/08), by the SemiSpinNet (Grant No. PITNGA-2008-215368), and by the Austrian FWF (P20065, P22477, P20550).

\appendix

\end{document}